\documentclass[prb,showpacs,twocolumn,preprintnumbers,amsmath,amssymb,floatfix]{revtex4}
\usepackage[dvips]{graphicx}
\usepackage[latin1]{inputenc}
\begin{document}
\draft

\newcommand{\lxpc} {Li$_{x}$ZnPc }
\newcommand{\lpc} {Li$_{0.5}$MnPc }
\newcommand{\etal} {{\it et al.} }
\newcommand{\ie} {{\it i.e.} }
\newcommand{\ip}{${\cal A}^2$ }

\hyphenation{a-long}

\title{Evidence for impurity-induced frustration in La$_2$CuO$_4$}

\author{P. Carretta$^{1}$, G. Prando$^{1,2}$, S. Sanna$^{1}$, R. De Renzi$^{3}$, C. Decorse$^{4}$  and P. Berthet$^{4}$}

\address{$^{1}$ Department of Physics ``A. Volta'', University of Pavia-CNISM, I-27100 Pavia, Italy}
\address{$^{2}$ Department of Physics ``E.Amaldi'', University of Roma Tre-CNISM, I-00146 Roma, Italy}
\address{$^{3}$ Department of Physics, University of Parma-CNISM, I-43124 Parma, Italy}
\address{$^{4}$ LPCES-ICMMO, Universit\'e Paris Sud, F-91405 Orsay Cedex, France}

\widetext

\begin{abstract}

Zero-field muon spin rotation and magnetization measurements were performed in La$_2$Cu$_{1-x}$M$_x$O$_4$, for
$0\leq x\leq 0.12$, where Cu$^{2+}$ is replaced either by M=Zn$^{2+}$ or by M=Mg$^{2+}$ spinless impurity. It is
shown that while the doping dependence of the sublattice magnetization ($M(x)$) is nearly the same for both
compounds, the N\'eel temperature ($T_N(x)$) decreases unambiguously more rapidly in the Zn-doped compound. This
difference, not taken into account within a simple dilution model, is associated with the frustration induced by
the Zn$^{2+}$ impurity onto the Cu$^{2+}$ antiferromagnetic lattice. In fact, from $T_N(x)$ and $M(x)$ the spin
stiffness is derived and found to be reduced by Zn doping more significantly than expected within a dilution
model. The effect of the structural modifications induced by doping on the exchange coupling is also discussed.

\end{abstract}

\pacs {76.75.+i, 75.10.Jm, 74.72.-h} \maketitle

\narrowtext

La$_2$CuO$_4$ has been the subject of an intense research activity since after the discovery of high temperature
superconductivity (HT$_c$SC).\cite{Muller} In fact, besides being the parent of HT$_c$SC it has been early
realized that La$_2$CuO$_4$ is one of the best prototypes of two-dimensional S=1/2 Heisenberg antiferromagnet on a
square lattice (2DQHAFSL).\cite{CHN} Accordingly, an impressive amount of theoretical and experimental studies on
the effects of charge and spin doping in La$_2$CuO$_4$ have followed.\cite{Johnst,Alloul} In this respect a
significant interest has attracted Zn$^{2+}$ S=0 for Cu$^{2+}$ S=1/2 substitution, which is known to be one of the
most detrimental for HT$_c$SC.\cite{ZnTc,Alloul} It has been observed that in La$_2$CuO$_4$ the substitution of
Cu$^{2+}$ with a spinless impurity can still be basically described by the 2DQHAFSL Hamiltonian, with a spin
stiffness ($\rho_s$) renormalized by the spin dilution.\cite{Cheong,PC1,PC2,Vajk,Neto} Namely, for
La$_2$Cu$_{1-x}$Zn$_x$O$_4$ one has
\begin{equation}
\label{H1} \mathcal{H}= - J \sum_{i,j} P_{\infty}(x) \mathbf{S_i}. P_{\infty}(x)\mathbf{S_j}
  \;\;\; ,
\end{equation}
where $J$ is the superexchange coupling among the nearest neighbour (n.n.) spins and $P_{\infty}(x)$ is the
probability to find a spin at site $i$ or $j$ for a doping level $x$, which for $x\rightarrow 0$ can be
approximated to $(1-x)$. Then, the above Hamiltonian can be mapped onto a 2DQHAFSL Hamiltonian, with
$2\pi\rho_s(x)= 1.15 J (1-x)^2$.\cite{CHN,PC1} This simple dilution model is able to describe to a reasonable
extent the basic properties of La$_2$CuO$_4$ doped with S=0 impurities and, in particular, the main features of
the zero-temperature sublattice magnetization $M(x)$ and of the N\'eel temperature $T_N(x)$.\cite{PC1,PC2,Vajk}
Nevertheless, more recent accurate studies\cite{Sasha2,Gingras} have clearly pointed out that $M(x)$ and $T_N(x)$
experimental data for La$_2$Cu$_{1-x}$Zn$_x$O$_4$  do not agree with the most accurate description of a 2DQHAFSL
achievable through numerical simulations. It is concluded that, although the dilution model is a good starting
point to analyze the properties of La$_2$Cu$_{1-x}$Zn$_x$O$_4$, a complete understanding of the effect of a
spinless impurity in La$_2$CuO$_4$ is still missing. Two years ago Liu et al.\cite{Sasha3} have suggested that S=0
impurities might induce magnetic frustration. In fact, if the energy of the lowest unoccupied impurity orbitals is
close to the one of Cu$^{2+}$ 3$d_{x^2-y^2}$ orbital, sizeable next n.n. superexchange couplings may arise and
compete with the n.n. coupling.\cite{Sasha3} Accordingly, a reduction of $\rho_s(x)$ and of $T_N(x)$ faster than
expected according to a dilution model is envisaged.

\begin{figure}[h!]
\vspace{5cm} \includegraphics{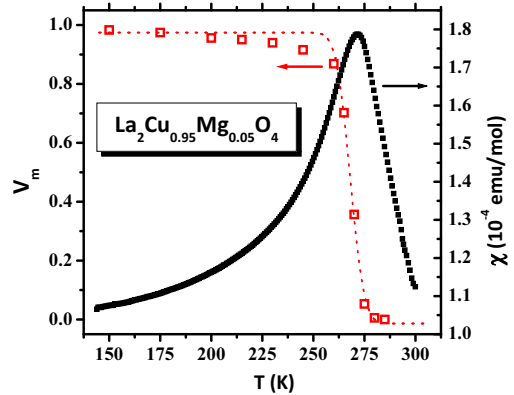} \caption{\label{Fig1} Typical
temperature dependence of the macroscopic static uniform spin susceptibility (closed squares, right vertical
scale) and of the magnetic volume fraction as derived from $\mu$SR measurements (open squares, left vertical
scale) in La$_2$Cu$_{1-x}$Mg$_{x}$O$_4$. }
\end{figure}

In order to test if impurity-induced frustration is really relevant we have performed a systematic comparison of
$M(x)$ and $T_N(x)$, determined by muon spin rotation ($\mu$SR) and magnetization measurements, in
La$_2$Cu$_{1-x}$Zn$_x$O$_4$ and in La$_2$Cu$_{1-x}$Mg$_x$O$_4$, for equal doping levels in the range $0\leq x\leq
0.12$. Since the energy of the unoccupied Mg$^{2+}$ orbitals is far away from the one of Cu$^{2+}$ 3$d_{x^2-y^2}$
orbital, no competing superexchange coupling should be present for Mg-doping, while for Zn-doping sizeable next
n.n. superexchange couplings may arise. It is found that La$_2$Cu$_{1-x}$Mg$_x$O$_4$ can be described within a
dilution model provided that minor modifications of $J$ due to the structural deformations are considered. On the
other hand, in La$_2$Cu$_{1-x}$Zn$_x$O$_4$ a reduction of $\rho_s(x)$ faster than expected for a dilution model is
found. We argue that this decrease of $\rho_s(x)$ indicates that $S=0$ impurities induce frustration.\cite{Sasha3}
It is also pointed out that for Zn doping the behaviour of $M(x)$  cannot be simply described by a 2DQHAFSL
Hamiltonian, since the geometry of the underlying magnetic lattice is progressively changing from the square
lattice one to the one of a diluted J$_1$-J$_2$ model\cite{Papin} as the doping increases.

\begin{figure}[t!]
\vspace{8.1cm} \includegraphics{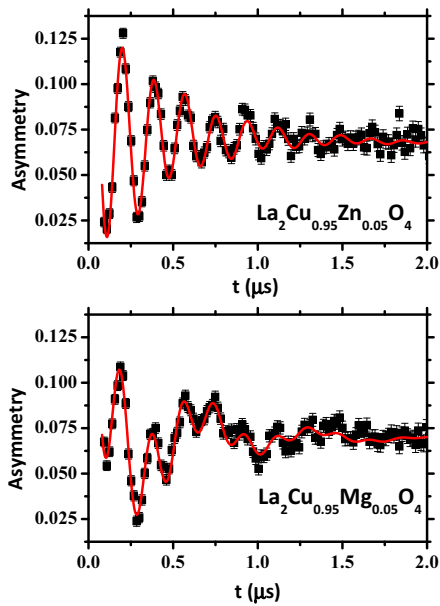} \caption{\label{Fig2} Time
evolution of the zero-field muon asymmetry for the $x=0.05$ Zn-doped (top) and Mg-doped (bottom) samples, at $T=
50$ K. The solid lines are the best fits according to Eq. 2 in the text.}
\end{figure}

Polycrystalline samples of La$_2$Cu$_{1-x}$M$_x$O$_4$ (M=Mg or Zn, x=0.02, 0.05, 0.075, 0.1 and 0.12) have been
obtained by standard solid state reaction of 99.99\% pure CuO, La$_2$O$_3$ and MgO or ZnO. Stoichiometric
quantities of these oxides were thoroughly mixed, pressed into pellets and multiple sinterings (12 to 24 h), with
intermediate grindings, were performed under air, in the temperature ($T$) range 900-1100°C. In order to avoid
oxygen excess, a last thermal treatment of 12h was performed under Ar gas flow at 800°C. X-ray diffraction showed
that the samples were single phase. Rietveld analysis was then performed using the space group \textit{Bmab} and
the lattice parameters were obtained.  The $T$-dependence of the static uniform spin susceptibility, determined
with a superconducting quantum interference device magnetometer, is characterized by a peak which marks $T_N$
(Fig. 1). The sharpness of the peak confirms the correct oxygen stoichiometry.\cite{Sharp}

Zero-field (ZF) $\mu$SR measurements were performed at ISIS pulsed muon source  on the MUSR beam line. Below $T_N$
the ZF muon polarization was characterized by clear oscillations associated with the onset of the long-range
magnetic order (Fig.2) and by the correspondent decrease in the longitudinal component of the muon
asymmetry.\cite{Blundell} It was noticed that, particularly for La$_2$Cu$_{1-x}$Mg$_x$O$_4$, two muon precessional
frequencies are evident (Fig.2), indicating two different muon sites. Thus, the decay of the ZF muon asymmetry
could be nicely fit to \cite{Blundell}
\begin{eqnarray}
\label{ZFAsymm}
      A(t)= A_1 e^{-\sigma_1 t} cos(\gamma_{\mu}B_1^{\mu}t + \phi) + \nonumber\\
      + A_2 e^{-\sigma_2 t} cos(\gamma_{\mu}B_2^{\mu}t + \phi) +
      A_3 e^{-\lambda t} + Bck
  \;\; ,
\end{eqnarray}
where $\gamma_{\mu}$ is the muon gyromagnetic ratio, $B_{1,2}^{\mu}$ is the local field at the muon sites 1 or 2,
which are characterized by different dipolar hyperfine couplings $D_i$ yielding $B^{i}_{\mu}=
D_i\langle\mathbf{S}\rangle$, with $\langle\mathbf{S}\rangle$ proportional to the sublattice magnetization.
$\sigma_{1,2}$ are the decay rates of the oscillating components, which increase with doping owing to the
enhancement of the local field distribution with increasing disorder. $Bck$ is a small constant term arising from
muons stopping in the sample environment and $\lambda$ is the decay rate of the muon longitudinal polarization.
While in La$_2$Cu$_{1-x}$Zn$_x$O$_4$ $A_1\gg A_2$, in the Mg-doped system the amplitude of the two components is
similar (Fig.2). \textit{The different ratio $A_2/A_1$ found for the two systems indicates that, in spite of the
same ionic charges and of the nearly equal ionic radii of Zn$^{2+}$ and Mg$^{2+}$, the lattice potential around
the two types of impurity is different.} From the drop in the longitudinal component $A_3$ of the muon asymmetry
below $T_N$ it is possible to derive the magnetic volume fraction $V_m$, since for a powder sample
$V_m=(3/2)[1-(A_3/(A_1+A_2+A_3))]$.\cite{Blundell} One observes a sharp increase in $V_m$ at $T_N$ (Fig.1). In
Fig.3 the $T$-dependence of $B_{1}^{\mu}$, normalized to its value extrapolated to $T\rightarrow 0$
$B_{1}^{\mu}(0)$, is reported as a function of $T/T_N$. One notices that, no matter which is the doping level, the
$T$-dependence of the local field at the muon is the same once the two aforementioned quantities are properly
rescaled.

\begin{figure}[h!]
\vspace{9.5cm} \includegraphics{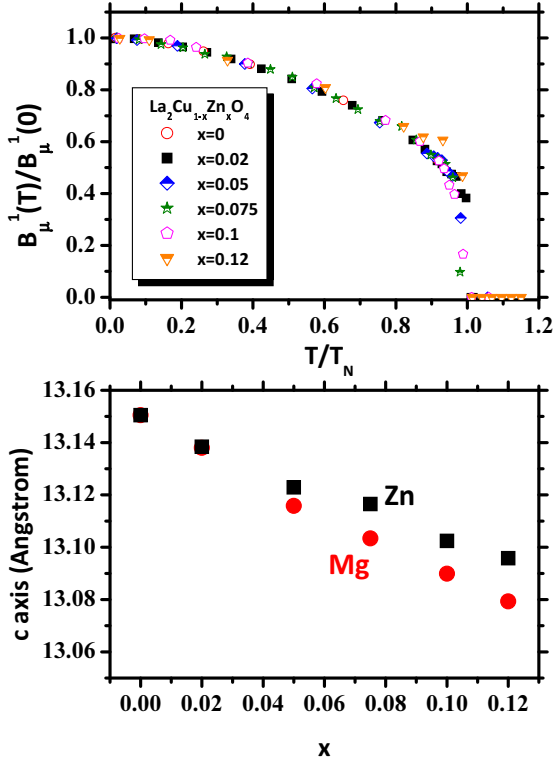} \caption{\label{Fig3} (Top) The
local field at the muon site 1, rescaled by its value for $T\rightarrow 0$, is reported as a function of
$T/T_N(x)$ in La$_2$Cu$_{1-x}$Zn$_x$O$_4$. (Bottom) Doping dependence of the c-axis length in
La$_2$Cu$_{1-x}$Mg$_x$O$_4$ and in La$_2$Cu$_{1-x}$Zn$_x$O$_4$.}
\end{figure}

In Fig.4 $T_N(x)$ and $M(x)/M(0)$, derived from the ratio $B_{1}^{\mu}(x,T\rightarrow
0)/B_{1}^{\mu}(0,T\rightarrow 0)$, are reported for the two systems. First of all, one notices that Zn doping
induces a more rapid decrease of $T_N$ than Mg substitution. In both cases $T_N(x)$ shows a linear decrease
described by $T_N(x)/T_N(0)= (1 - \alpha x)$, with an initial slope $\alpha_{Zn}= 3.52\pm 0.17$ for Zn doping and
$\alpha_{Mg}= 2.7\pm 0.15$ for Mg doping. Given the energy separation between the first excited Mg$^{2+}$ orbitals
and Cu$^{2+}$ $3d$ orbitals, the latter $\alpha$ value should be taken as the one expected for a diluted
2DQHAFSL\cite{Sasha2} ($\alpha= 3.196$), reduced by the inter-plane coupling $J_{\perp}$.\cite{lattice} However,
to properly analyze those results, one has first to consider if the lattice deformation induced by doping may
cause a significant variation in the n.n. coupling $J$.\cite{lattice} Namely if $J$ in Eq. 1 depends on doping.
First of all we point out that the changes in the lattice parameters show a similar trend for both types of
substitutions (Fig. 3). One notices also that the $c$ axis changes less in the Zn doped compound, where $T_N$
decreases faster. Moreover, even in the x=0.05 samples where the difference in La$_2$Cu$_{1-x}$Zn$_x$O$_4$ and
La$_2$Cu$_{1-x}$Mg$_x$O$_4$ lattice parameters is less than 1/1000 a significant difference in the N\'eel
temperatures is observed. These latter observations show that \textit{the major differences in the $T_N(x)$ of the
two compounds should not be associated with the effect of the lattice distortions on $J(x)$ } \cite{lattice} but
rather, as will be discussed later on, with the frustration induced by Zn-doping.

\begin{figure}[t!]
\vspace{12cm} \includegraphics{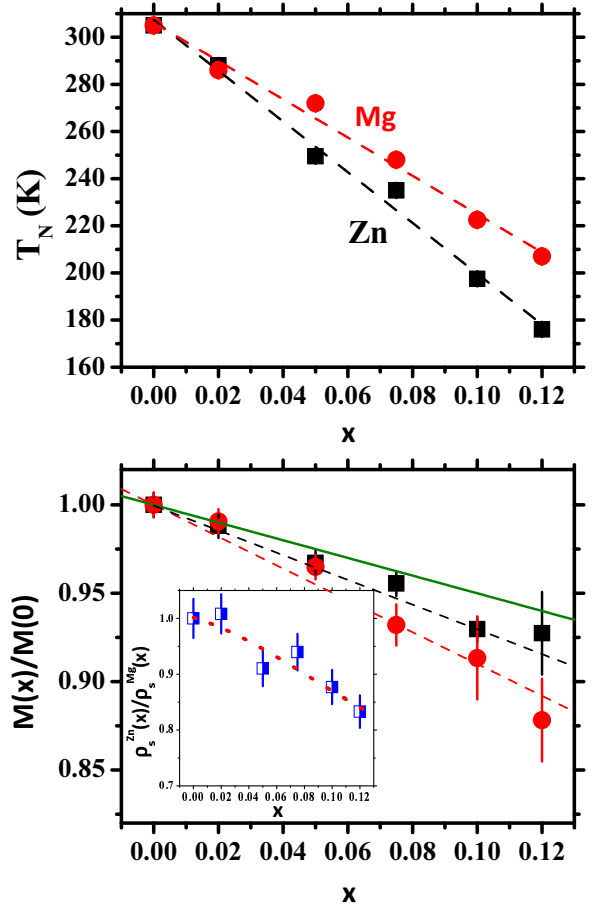} \caption{\label{Fig4}  Doping
dependence of $T_N(x)$ (top) and of $M(x)/M(0)=B^{1}_{\mu}(x,T\rightarrow 0)/B^{1}_{\mu}(0,T\rightarrow 0)$
(bottom) in La$_2$Cu$_{1-x}$Mg$_x$O$_4$ and in La$_2$Cu$_{1-x}$Zn$_x$O$_4$ ($B^{1}_{\mu}(0,T\rightarrow 0)= 427$
Gauss). The dashed lines are guides to the eye, while the solid line in the bottom panel shows the behaviour
expected for a diluted QJ1J2SL (see text). In the inset the ratio of the spin stiffness derived for Zn and
Mg-doped compounds is reported.}
\end{figure}


One can estimate  the variation of $J(x)$ due to the lattice strains on the basis of the experimental data for
$T_N(x)$ and $M(x)$ reported in Fig.4. In fact, one can write\cite{CHN,PC1,Neto}
\begin{equation}
\label{EqTN} T_N(x)= J_{\perp} P_{\infty}(x) \biggl( \frac{M(x)}{M(0)}\biggr)^2 \xi^2(T_N,x)
  \;\;\; .
\end{equation}
Since $T_N\ll J$ one has $\xi(T_N,x)\sim exp(2\pi\rho_s(x)/T_N(x))$.\cite{CHN} Now, assuming that for
La$_2$Cu$_{1-x}$Mg$_x$O$_4$ a simple dilution model works one can write $2\pi\rho^{Mg}_s(x)= 1.15 J(x)(1-x)^2$
(see Eq.1 and Ref.\onlinecite{CHN}). For a fixed $x$ value one notices from Eq. \ref{EqTN} that an increase in
$J(x)$ causes a correspondent enhancement of $T_N(x)$. By taking the experimental values for $T_N(x)$ and $M(x)$
one derives an increase of $J(x)$ by only $\sim 40$ K for the x=0.12 sample,\cite{Perp} with respect to
$J(0)\simeq 1580$ K.\cite{Johnst} This increase is close to the one that could be estimated from the $c$-axis
contraction, by taking into account the c-axis compressibility\cite{Pressc} and the variation of $J$ with
pressure.\cite{Raman} Thus it is concluded that the lattice strains would affect $J(x)$ and $\alpha_{Mg}$ only to
a minor extent.


Now, we shall determine the variation of the spin-stiffness $\rho^{Zn}_s(x)$ due to Zn doping. Given the similar
trend of the lattice parameters in Mg and Zn doped samples (Fig.3), it is reasonable to use also for
La$_2$Cu$_{1-x}$Zn$_x$O$_4$ the values of $J(x)$ derived for La$_2$Cu$_{1-x}$Mg$_x$O$_4$. From Eq. \ref{EqTN}, by
taking the ratio of $T_N(x)$ for Zn and Mg-doped compounds one can write
\begin{equation}
\label{Eq3} \rho^{Zn}_s(x)= \frac{T^{Zn}_N(x)}{4\pi}ln\biggl( \frac{t(x)}{m^2(x)}\biggr) +  \frac{2.3
J(x)t(x)(1-x)^2}{4\pi}
  \;\;\; ,
\end{equation}
where $t(x)= T^{Zn}_N(x)/T^{Mg}_N(x)$, while $m(x)=M^{Zn}(x)/M^{Mg}(x)$. By taking the experimental data for
$T_N(x)$ and $M(x)$, on the basis of Eq.\ref{Eq3}, one finds a systematic reduction of $\rho^{Zn}_s(x)$ with
respect to $\rho^{Mg}_s(x)$, which reaches about 16\% in the x=0.12 sample (inset to Fig. 4). Namely, there is a
decrease of the spin-stiffness for Zn-doped samples which is much faster than the one expected for a dilution
model. Since the reduction of $\rho^{Zn}_s(x)$ cannot be due to lattice strains, as mentioned above, it is clear
that the observed variation should be associated with a different effect of Zn and Mg orbitals on the
superexchange coupling mechanisms. In particular \textit{the marked reduction of the spin stiffness reported in
Fig.4 suggests that Zn doping gives rise to interactions that compete with $J$}, namely that there is a
frustration induced by the impurities, consistently with previous theoretical predictions.\cite{Sasha3}

If now one considers the doping dependence of $M(x)$ one finds a very similar trend both for Mg and Zn-doped
samples (Fig.4). Again, one can consider that La$_2$Cu$_{1-x}$Mg$_x$O$_4$ behaves as a diluted spin system with
$J(x)$ slightly increasing with doping due to the structural modifications. Also here it is instructive to observe
that the lattice contraction associated with the application of an external pressure yields a small but
non-negligible reduction of $M(x)$.\cite{PressM} Thus, it is tempting to associate the slight decrease of $M(x)$
found in La$_2$Cu$_{1-x}$Mg$_x$O$_4$ with respect to the dilution model\cite{Sasha2} to lattice effects. On the
other hand, it is not clear why in La$_2$Cu$_{1-x}$Zn$_x$O$_4$, where $\rho^{Zn}_s(x)$ is reduced, $M(x)$ does not
show a more pronounced decrease. This implies that the more significant reduction of $T_N(x)$ found in
La$_2$Cu$_{1-x}$Zn$_x$O$_4$ originates from the decrease of $\rho^{Zn}_s(x)$ rather than from the one of $M(x)$
(see Eq. 3).

In order to understand why $M(x)$ reduction is slightly lower in La$_2$Cu$_{1-x}$Zn$_x$O$_4$ one has to consider
that if Zn is giving rise to a next n.n. coupling, the magnetic lattice around the impurity is no longer a square
lattice\cite{Sasha3} and the mapping of the microscopic Hamiltonian onto a 2DQHAFSL Hamiltonian should no longer
be valid. Indeed, in that case the local configuration around Zn would be more similar to the one found in the
diluted $S=1/2$ $J_1-J_2$ model on a square lattice (QJ1J2SL),\cite{Papin} where the more connected spin texture
yields a less pronounced decrease of $M(x)$. In fact, in Li$_2$V$_{1-x}$Ti$_x$SiO$_4$,\cite{Papin} a prototype for
the diluted QJ1J2SL model, the initial decrease of $M(x)/M(0)\simeq 1 - 0.46 x$  is  lower than the one found for
a diluted 2DQHAFSL (Fig. 4).\cite{Sasha2} Still one has to consider the effect of $\rho^{Zn}_s(x)$ reduction on
$M(x)$. In the framework of the $J_1-J_2$ model\cite{Richter} the observed reduction of $\rho^{Zn}_s(x)$ would
correspond to an effective increase of the ratio $J_2/J_1$ to$ \simeq 0.04$ for $x=0.12$, which should lead to a
correspondent reduction of $M(x)/M(0)$ by only 3.6\%,\cite{Richter} i.e. to an additional decrease
$M(x)/M(0)\simeq 1 - 0.3 x$. Thus, taking into account both the effect of dilution and the increase of frustration
the observed decrease of $M(x)/M(0)\simeq 1 - 0.75 x$ due to Zn doping (Fig. 4) seems reasonable. Qualitatively
speaking, the microscopic configuration that one should find in La$_2$Cu$_{1-x}$Zn$_x$O$_4$ is in between the one
of a diluted 2DQHAFSL and the diluted QJ1J2SL model and, accordingly, $M(x)/M(0)$ curve should stay in between the
trend expected for those two models (Fig.4), as it is observed.

In conclusion, from the comparison of $M(x)$ and $T_N(x)$ in Mg and in Zn doped La$_2$CuO$_4$, it was found that
La$_2$Cu$_{1-x}$Mg$_x$O$_4$ can still be described in terms of a dilution model with minor corrections due to
lattice strains. On the other hand, the marked reduction of the spin stiffness found in
La$_2$Cu$_{1-x}$Zn$_x$O$_4$ indicates that in this latter system competing next n.n. interactions may arise around
the impurity and generate a frustrated magnetic lattice. Also the reduced dependence of $M(x)$ found in the
Zn-doped system suggests the presence of a spin texture more connected than the one of a diluted 2DQHAFSL, which
indicates a different effect of Zn and Mg orbitals in the superexchange coupling mechanisms.


Useful discussions with A.L.Chernyshev and the technical assistance by Sean Giblin during the measurements at ISIS
are gratefully acknowledged.




\begin{references}
\bibitem{Perp} The variation of $J_{\perp}$ is neglected since, as it is shown in Ref.\onlinecite{Raman}, the
lattice strain induced by pressure affects mainly the in-plane coupling rather than $J_{\perp}$.
\end{references}
\end{document}